# THE COMMISSIONING PHASE •

F. Bortoletto[1], S. Benetti[1], G. Bonanno[2], C. Bonoli[3], P. Bruno[2], C. Carmona[1], P. Conconi[4], L. Corcione[5], R. Cosentino[1], M. D'Alessandro[3], R. Dominguez[1], D. Fantinel[3], A. Galli[1], D. Gardiol[1], A. Ghedina[1], F. Ghinassi[1], E. Giro[3], C. Gonzales[1], NI. Gonzalez[1], J. Guerra[1], A. Magazzu'[1], D. Mancini[6], E. Marchetti[1], J. Medina[1], F. Pasian[7], F. Paulli[1], C. Pernechele[3], M. Pucillo[7], R. Ragazzoni[3], C. Riverol[1], L. Riveroi[1], P. Schipani[6], R. Smareglia[7], G. Tessicini[1], G. Trancho[1], C. Vuerli[7], A. Zacchei[7]

[1] *Centro Galileo Galilei - TNG, La Palma, Canarie (E)*
[2] *Osservatorio Astrofisico di Catania*
[3] *Osservatorio Astronomico di Padova*
[4] *Osservatorio Astronomico di Milano*
[5] *Osservatorio Astronomico di Torino*
[6] *0sservatorio Astronomico di Napoli*
[7] *Ossenwtorio Astronomico di Trieste*

## 1. Introduction

In May 1997, a consistent part of the services and structures committed to the industry had already been released to the *commissioning* group. The telescope itself was, with the exception of the Nasmyth derotators, motors and all the optics groups, basically ready in its mechanical parts to accept the integration of all services and control equipment. Also the verification of the cabling (interlocks, data-nets, power and controls) already mounted was started in the same period.

Considering the telescope alone, the work made during the commissioning phase has been subdivided into the following parts:

- Telescope cabling (nets, controls, interlocks etc.)
- Servo system mounting and tuning
- Optics mounting and alignments
- Informatics
- Telescope tests and tuning

Starting from June 1998 (telescope first-light date) the telescope went gradually in use, several nights per week, in order to test and tune the tracking and pointing system, the optics and the first derotator system (Nasmyth A station). At the end of the commissioning period and with the first scientific instruments mounted (April 1999) also the first routinely observations started. In this moment the telescope is doing astronomy 80% of time and the complete first-light instrumentation is mounted.

The overall group of persons involved, partially or in full-time, in the *commissioning* phase is shown in Tab.1.1. The group was also completed with two secretaries (P. Mazzucato and N. L. Jorge Martin), operating respectively in Padova and La Palma, and with an administrative responsible at the *Wilda del Progetto Galileo* (UPG) in Padova (S. Cestaro).

In the first three columns we list the personnel in charge at the *Centro Galileo Galllei* (CGG) based in La Palma, who were active during the commissioning phase, and other personnel contracted by the *Consorzio Nazionale per l'Astronomia e l'Astrofisica* (CNAA), while the other mentioned people are from Italian observatories. It is worth to mention that none of the people in the CGG staff had previous working experience at the TNG during the construction phase.

Tab. 1.1 - The TNG **commissioning group**

| CGG | CGG | UPG | OAPD | OAC | OACT |
|---|---|---|---|---|---|
| F. Bortoletto | C. Camiona | N. Boaretto | C. Bonoli | E. Cascone | G. Bonanno |
| S. Benetti | R. DeAngelis | G. Canton | L. Contri | D. Mancini | P. Bruno |
| D. Gardiol | A. Galli | G. Giudici | M. D`Alessandro | P. Schipani | |
| A. Ghedina | C. Gonzales | L. Oculi | D. Fantinel | | |
| F. Ghinassi | J. Medina | | E. Giro | | |
| A. MagazzU | G. Piselli | | C. Pernechele | | |
| E. Marchetti | C. Riverol | | R. Ragazzoni | OATO | OATS |
| G. Tessicini | L. Riverol | | Strazzabosco | L. Corcione | F. Pasian |
| A. Zacchei | F. Rossi | | L. Traverso | | R. Smareglia |
| F. Paulli | M. Gonzalez | | A. Frigo | | C. Vuerli |
| G. Trancho | R. Dominguez | | G. Farisato | | |
| .1. Guerra | R. Cosentino | | | | |

Notes. CGG:Centro Galileo Galilci, La Palma: UPG:Ufficio del Progetto Galileo: Astronomical Observatories:
OAPD = Padova; OAC = Capodimonte (NA); OACT = Catania: OATO = Torino; OATS = Trieste

With the beginning of routinely astronomical operations, the group changed with an increasing number of people dedicated to nocturnal operations and to astronomers' support, while some of the former staff technicians did stay on.

## 2. The schedule

The schedule of the commissioning work is shown in Tab.2.1; it reports the progress of the principal activities on the telescope, the implementation of the first light instruments and the first observational phases. The blue marks on the left side of the table show the end of activities as foreseen in December 1996, after the telescope inauguration; in that case the estimated termination of activities and the start of the first experimental observations were placed around the end of 1997 (see the vertical blue bar). Red marks show the end of activities, as really happened; simply comparing the foreseen and expected dates, one can see a delay of about 16 months.

The main milestones arc:

- the telescope technical first-light in June-July 1998;

- the first-light of the derotator A (tracking and Shack-Hartmann cameras) in Sept-Oct 1998;

- the scientific instrumentation first-light with the visual imaging camera) in Nov 1998.

Since the delivery of the telescope to the commissioning group at the beginning of 1997 and during the first six months of activity, the group has dedicated a considerable part of time installing the structures and services (mechanical workshop, electronic lab, computers and network installations, basic safety) needed for the assembling of the optics train and the installation of the electrical systems on the telescope.

The real work on the telescope structure, in parallel with other major operations like drive system, service controls and interfacing with the computer system started during July 1997. Some negative impact in the schedule up to Feb 1998 was experienced with several acceptance and guarantee interventions, in particular:

- building cover, because of problems with rain and ice;

- hydrostatic-bearing system, due to problems with the smoothness of the sliding planes;

- cooling system, setting, failures and complete rebuild of the computer control system.

Tab. 2,1 TNG commissioning schedule during 1997-99

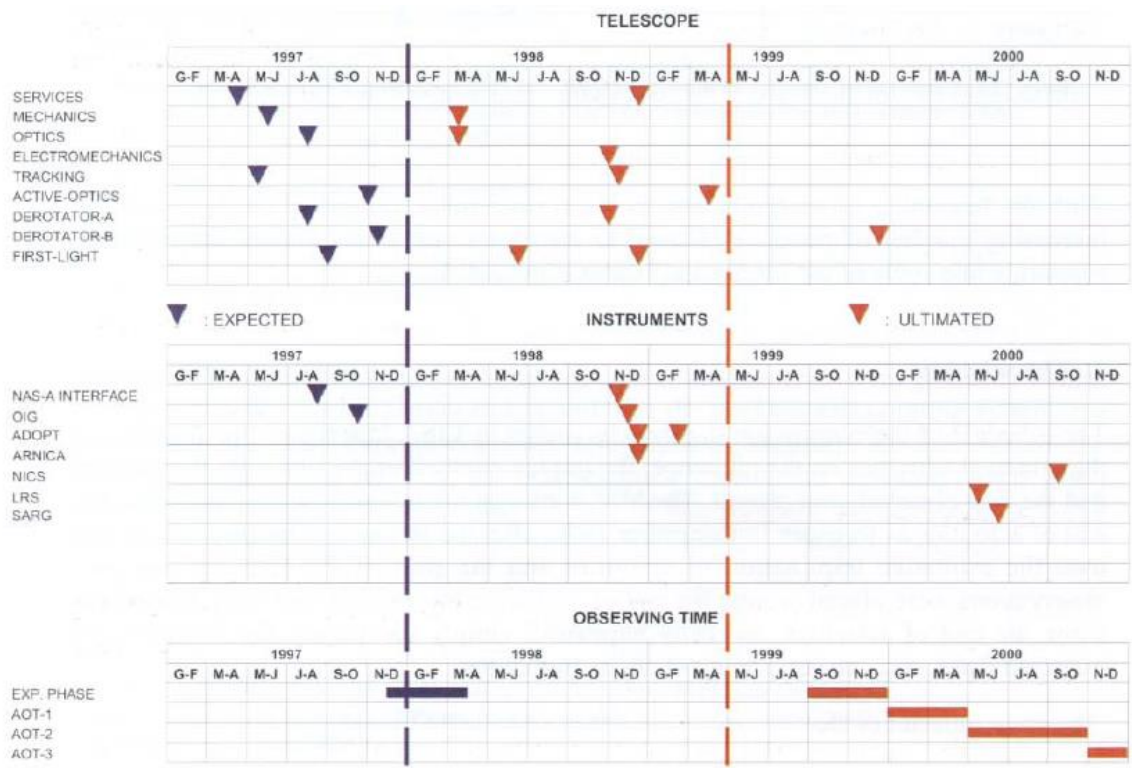

From a pure *planning* point of view, the main problem for the commissioning group in trying to fit a predefined schedule of work has been represented by the lack of a true pre-assembling and verification phase on several parts of the telescope, in particular:

- secondary and tertiary mirror-supports not pre-mounted at the telescope;
- secondary and tertiary mirrors not pre-mounted inside the mirror-frames;
- mirror mounting tools not verified;
- interlocks not mounted and tested;
- derotators never pre-mounted at the telescope and complete lack of the concerned electrical cabling;
- telescope cabling and telescope auxiliary services not tested on the field with the hardware and/or software definitive control.

Of particular relevance is the last point in that it implies that a consistent portion of the application software has been developed directly in site. often with the introduction of several arrangements *ad-hoc* not foreseen in the original plan. The latter has been, in some way, partially compensated by a better cooperation, in site, between the commissioning and the SW implementation groups.

As a conclusion in the schedule here shown there is also a consistent portion of construction and integration operations not part of a true *commissioning* activity.

Telescope *first-light,* as shown in Tab. 2.1. has been divided into four basic steps, each one necessary to verify the correct operation and performance of the system (in particular active optics and tracking) at different stages of complexity :

1. First light Phase-1, making use of a camera looking at the inner&outer focal planes;
2. First light Phase-2, making use of a Shack-Hartmann test camera (Antares);
3. First light Phase-3, making use of Derotator-A Shack-Hartmann and Tracking cameras.;
4. First light Phase-4 making use of the Optical Imager Galileo (OIG).

**3. The optics system**

TI G is an ALT-AZ telescope with two f-11 Nasmyth focal stations; it is based on a 3.5 meters thin and active primary mirror. The mounting of the three optics groups (M1, M2 and M3 mirrors plus supports) composing the aplanatic optical system of the telescope has been made in the following basic steps:

- the insertion of mirrors inside mirror frames (M2 and M3);

- the assembling of mirror groups inside templates (M2 and M3);

- the mechanical alignment of the telescope axis;

- the insertion of mirrors inside the telescope (M2, M1 and M3, in sequence);

- the definition of the telescope Optical Axis (OPT);

- the optical alignment of mirrors inside supports with the OPT axis.

Basically we followed the alignment steps reported in the ESO-NTT procedures, although major changes have been done on the mirror-blank mounting procedures and in the cross-alignment between the AZ and EL axes. It is to be noticed that the M1 mirror was already checked inside the active-cell in 93 in Zeiss, although never moved in the EL plane. Table 3.1 shows the numerical results taken after the mechanical and optical alignment phase summarized above, while Fig. 3.2 shows the same numbers reported in a sketch of the telescope main axes.

The precision in the finally obtained results shown in Tab. 3.1, is basically limited by the environmental seeing ($\pm$ 10 arcsec on the estimation of several angles). Nevertheless the precision is good because better than the thermal and mechanical effects typically experimented by the overall telescope structure when in operation. More details on the alignment phase are given in Ref. **1.**

**Tab. 3.1 - Final results of the mechanical and optical alignment**

|  | Tilt (") | Err (") | Direction ($^0$) | Decenter ($\mu$m) | Err, ($\mu$m) | Direction (°) |
|---|---|---|---|---|---|---|
| M1 Cell |  |  |  | 430 |  | +116 |
| M2 Azimuth | 2.5 | +10 | +90 | 120 | ±310 | +55 |
| M3 Bearing | 140 | ±18 | +143 | 360 | ±100 | +16 |
| M3 Spider | 12 | ±18 | +130 | 60 | ±100 | +100 |
| Target Nas A | 13 | +2 |  | 0 |  |  |
| Target Nas B | 45 | ±7 |  | 0 |  |  |
| G-AZ $\perp$ (verticality) | 3 |  | 31 |  |  |  |
| ALT-AZ $\perp$ | 23 |  |  |  |  |  |
| ALT-AZ (coplanarity) |  |  |  | 230 |  |  |

## 4. The pointing and tracking system

The telescope drive system was completed in 1998; the last part mounted at the telescope was the Nasmyth-B derotator in 1999. The major problem found in the commissioning of this system was represented by the incomplete cabling of the emergency systems during the construction phase and, in particular, by the discovery that the hydrostatic bearing sliding stream was not inside the planarity specification. This problem was at the origin of the abnormal AZ torque required to the drive system and visible in the first diagram of Fig. 4.1.

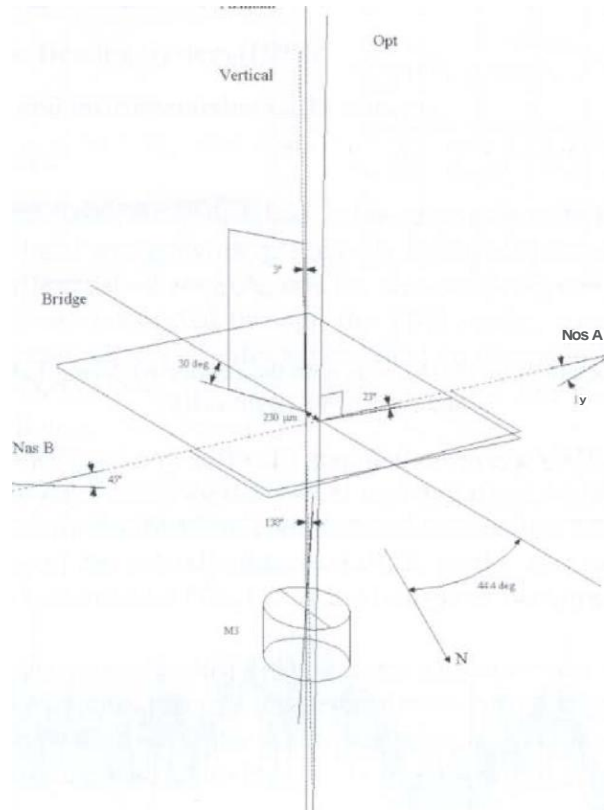

**Fig. 3.2: The main mechanical axes of the telescope found during the alignment at TNG**

Data were taken during Feb 1998 and they are the result of the friction of a local asperity in the sliding stream with each one of the 24 hydrostatic pads during an entire AZ revolution, at every contact there is a pique of torque with a period of 360/24 deg. The problem was resolved in March-April 1998 after a complicated and delicate operation of lapping made on the sliding stream and on every hydrostatic pad (see Ref. 5). The torque demand obtained after the operation is shown at the right side of the same figure and can be directly compared with the previous figure.

At the end of the whole intervention on the telescope hydrostatic bearing a final check on the tracking accuracy was performed. In Fig. 4.2 it is represented the response of the system to a step impulse of 0.1°: after an initial settling time, needed to reach the desired angle, the ISMS positioning error is of the order of 0.01 arcsec the test was made in the region affected by the previous problem.

## 5. The informatics system

The Informatics hardware configuration is based on a number of VME crates (7 up to now), dedicated to directly control the telescope subsystems and instrumentation and several HP workstations (about 8) for the remote user interface. Transputers and transputer links are dedicated to the active optics and the CCD camera control, some Personal Computer is also used for specific tasks needing graphical capabilities (the

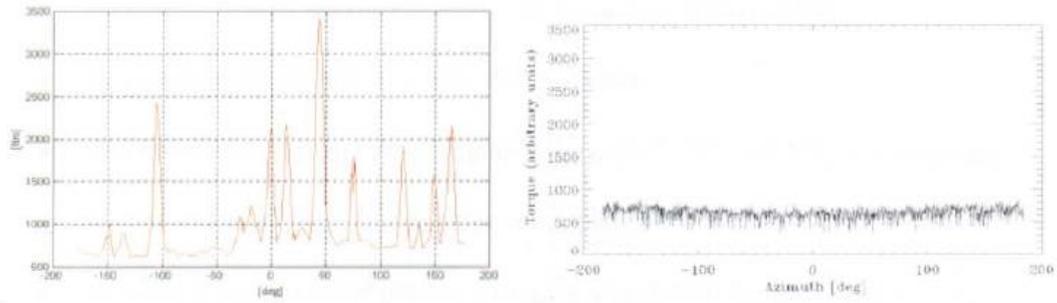

Fig. 4.1 — Torque behavior on the AZ axis with data collected during Feb 1998 (left side) and April 1998 (right side)

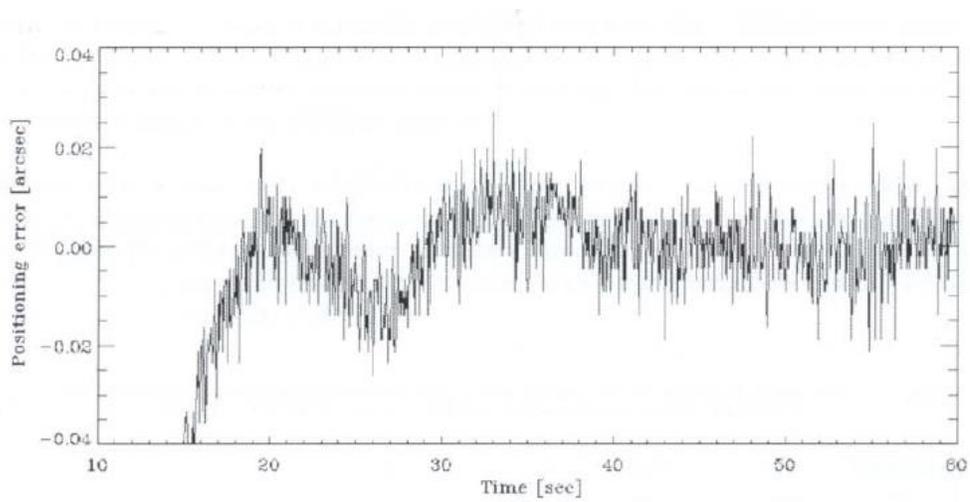

Fig. 4.2 — After-step positioning error @. AZ =44°

The subsystems under supervision are:

- Active optics;
- Telescope movements;
- UPS time services;
- Nasmyth A rotator (DERA);
- Nasmyth B rotator (DERB);
- Dome and telescope services (PLCs);

- Climatization system;
- Hydrostatic Bearing System (HBS);
- Ancillary and instrumentation CCD cameras;
- Data archives.

The network system, completely installed, is based on two main transport media: FDDI and Ethernet. The local area network is logically subdivided into three sub-nets, one for telescope and instrumentation controls, one for data archiving and one for public access. The three sub-nets are connected through the TNG router, equipped with two FDDI ports and two Ethernet ports. The software installed up to now is basically subdivided in three parts:

- **GATE**. Intermediate level software installed on every VME. It provides the basic functionality of each subsystem as: local user interface, local command execution, task creation, alarm notification, communication handling with the connected WSS and monitoring of the overall ambient. GATE works also as an interface to other heterogeneous systems like PCs, PLCs and transputer networks.

- **WSS**. High level software installed on every workstation in the telescope network. It handles the remote user interface communicating with the VME systems. Commands can be sent, telemetry is received and checked. It represents also the link between controls and data archiving.

- **IDL**. Algorithmic and graphical interlaces.

An example of user interface based on the IDL widget system is shown in Fig. 7.4, this is the case of the Shack-Hartmann frame analyzer.

## 6. The adapter derotators

The first adapter-derotator has been mounted during the month of Sept 1998 at the Nasmyth station A of the telescope, the same operation has been made for the B derotator at the end of 1999.

At present the derotator A is serviced by a CCD Peltier cooled camera covering the functions of Shack-Hartman analyzer (10x 10 and 40x40 sub-pupils) and tracker/fieldviewer. The second probe of the derotator is equipped with an intensified-CCD camera with the only function of field-viewer; a similar situation is present at the Nasmyth-B with the exception of the intensified camera substituted by a second CCD cooled camera.

The field covered by the camera in tracking-mode is more than 2'x2'; this can be seen on the [eft panel of Fig. 6.2, where a single one minute exposure of NGC752 taken during the first tests is shown. The stars-profile FWHM is of 1" confirming the good quality of the overall optics transport inside the derotator (six folding mirrors, a collimator plus a camera objective). In the same camera there is also provision of a standard photometric fitter set: the picture of Jupiter on the right panel of Fig, 6.2 was made during the same run, by making use of the B-V-R filter set.

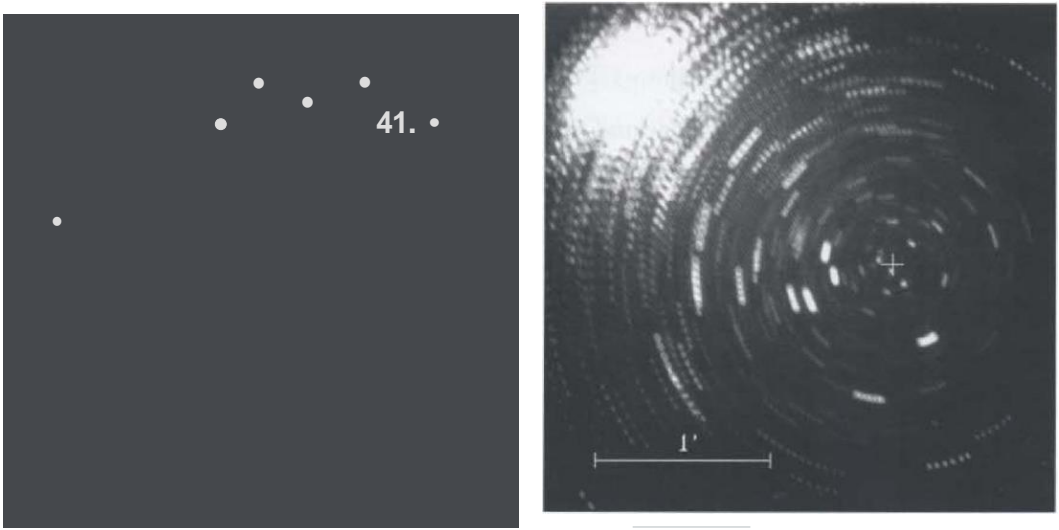

Fig. 6,1. - Finding the derotator mechanical center of rotation (left) and the parallatic rotation center (right)

A peculiar feature of TNG is that the same camera controllers are used both for technical and scientific cameras so also the technical frames (autoguider, image-analyzers) are FITS format files, archived and with good photometric quality (16 bits dynamics).

The calibration of the derotator was also started. In Fig. 6.1 the tests made in order to identify the *mechanical derotator axis* (left side) and the *optical parallactic axis* (right side) are shown. The first picture was obtained by co-adding three NGC 1039 exposures interleaved by a derotator manual rotation, the fit of the center being indicated by a cross. The second picture, taken with the derotator in steady position, was obtained by co-adding six M 15 exposures at different UT; the position of the parallactic center is again indicated by a cross and the fit with the previous measure is good.

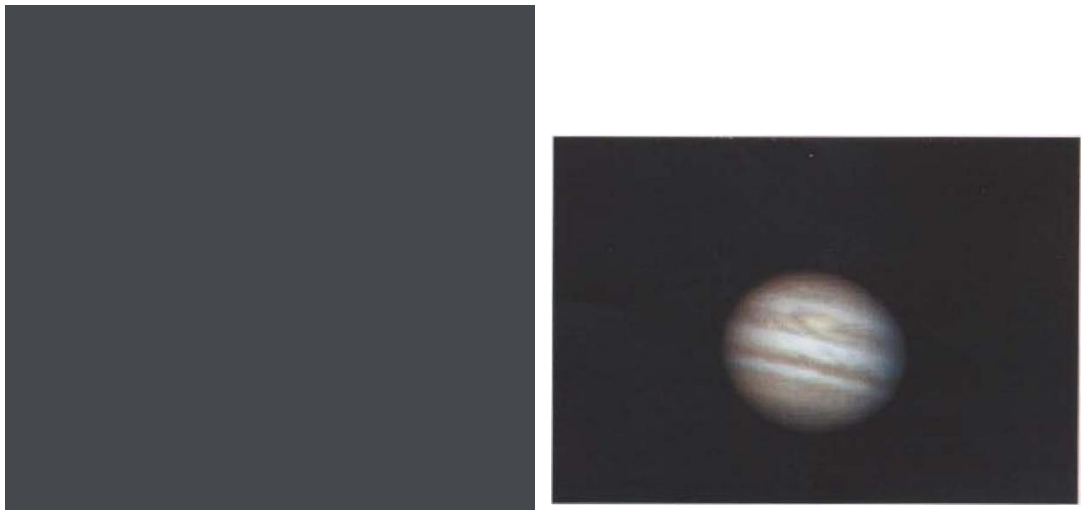

Fig. 6.2. — A single exposure through the derotator-A tracking camera (left), a three-chromatic visualization of Jupiter obtained from three exposures with filters B-V-R

## 7. The telescope First-Light

The telescope first-light was made during the course of June 1998. After a first look at the naked focal plane the check on the quality of the alignments was made looking at the *inner and outer focused* images of a single bright star (see also Ref 3) making use of one of the tracking cameras mounted on a movable carriage (see Fig. 7.1); also a check on the plate scale and the position of the focal plane was made.

The tv12 mirror was then commanded to move in such a way to correct its residual decentering after the alignment phase. After this correction the transverse coma aberration went down from ~0.59 arcsec to ~0.22 arcsec, while the spherical aberration was stable within about 0.10 arcsec.

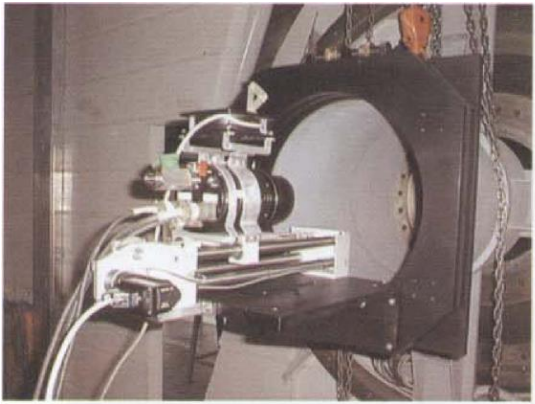
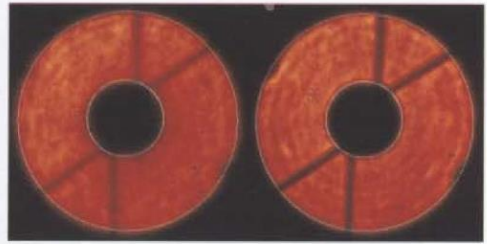

Fig. 7.1 — The camera truss at the naked arm-B focal plane and a couple of inner-outer pupil images used to evaluate the low-order optical distortions

The first object observed after the preliminary set-up was the double system *epsilon-lirae*. the result is shown in Fig. 7.2

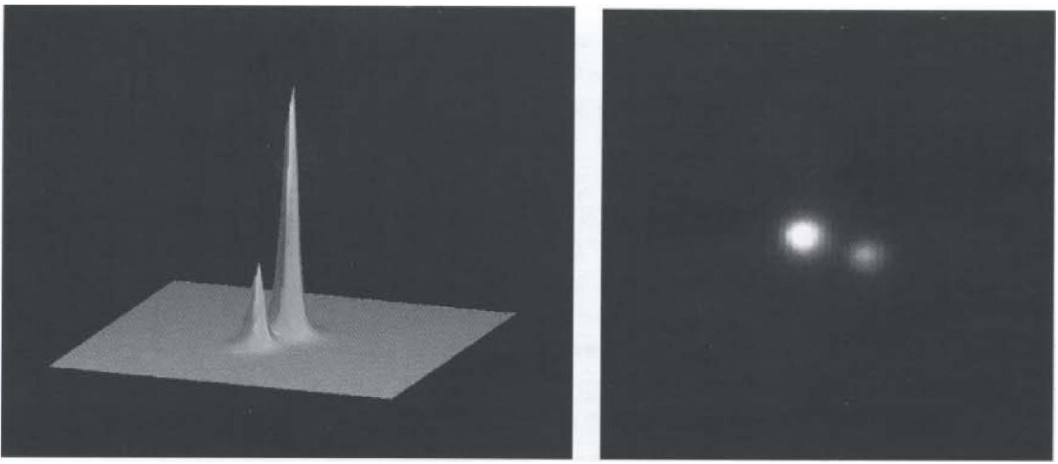

Fig. 7.2 - The first object observed: epsylon-lirae double-star system.
separation: 2.6", profile FWHM: 0.8"

A second example of exposure taken during the first-light is the globular cluster M92 in Hercules. In Fig. 7.3 a picture of the exposure is shown; the FWHM of the stars in the field is ~0.65 aresec. It is possible to notice a little drift along the *y-axis* of the chip, probably due to some residuals of tracking. The exposure time was 10 sec and the extension of the drift is about 0.85 arcsec. The seeing during the acquisition was 0.47 arcsec (IAC DIMM courtesy).

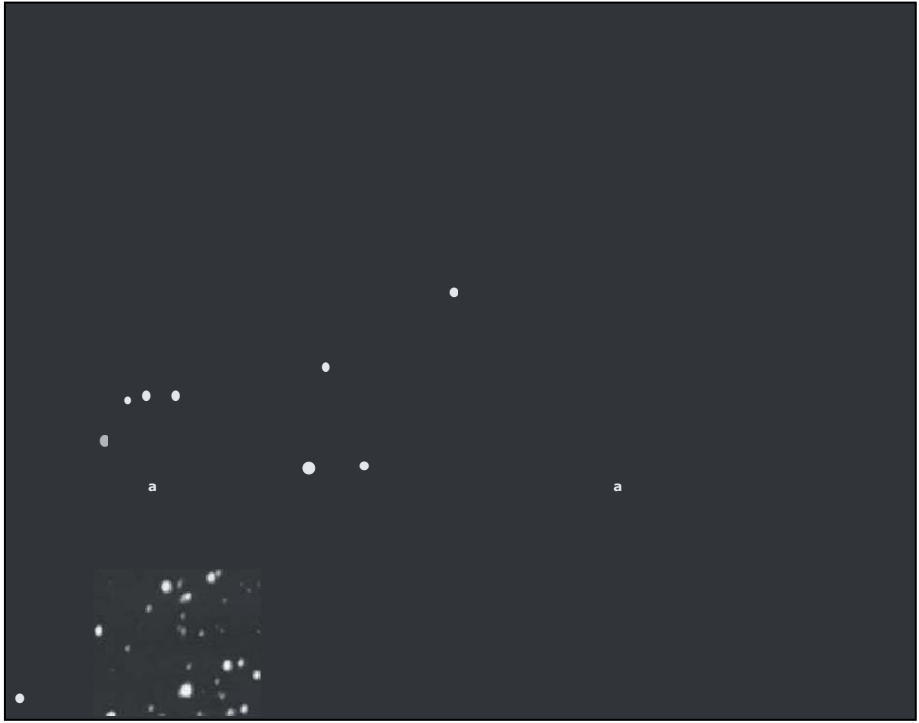

Fig. 7.3 — The M92 globular cluster in Hercules

After the first-light several tests on the active-optics system of TNG were made using an off-line Shack-Hartmann analyzer *(antares)* and the analyzers mounted into the derotators. Both the primary and the secondary systems were characterized during such tests.

A typical result from the first Shack-Hartmann (S-H) analysis runs (August $6^{1h}$, 1998) is shown in Fig. 7.4. On the right window of the background image the pupil S-H spots are shown jointly with their computed centers (red squares), while, on the left side, the same spots are shown superimposed to those spots found in the reference wavefront. The small window shows the results of the analysis: for every mode one can see the RMS wavefront distortion (nanometers plus the azimuthal angles) and the equivalent aberration introduced on the focal plane (aresec).

The values of the spherical (0.13 arcsec) and coma (0.2 arcsec) aberrations are comparable with those found in the precedent *pupil-analysis* (see "TNG first light report" document), I.E. about 0.1 and 0.22 arcsec respectively. In the results window the analytical reconstructed wavefront (without tilt and defocus modes) is also shown.

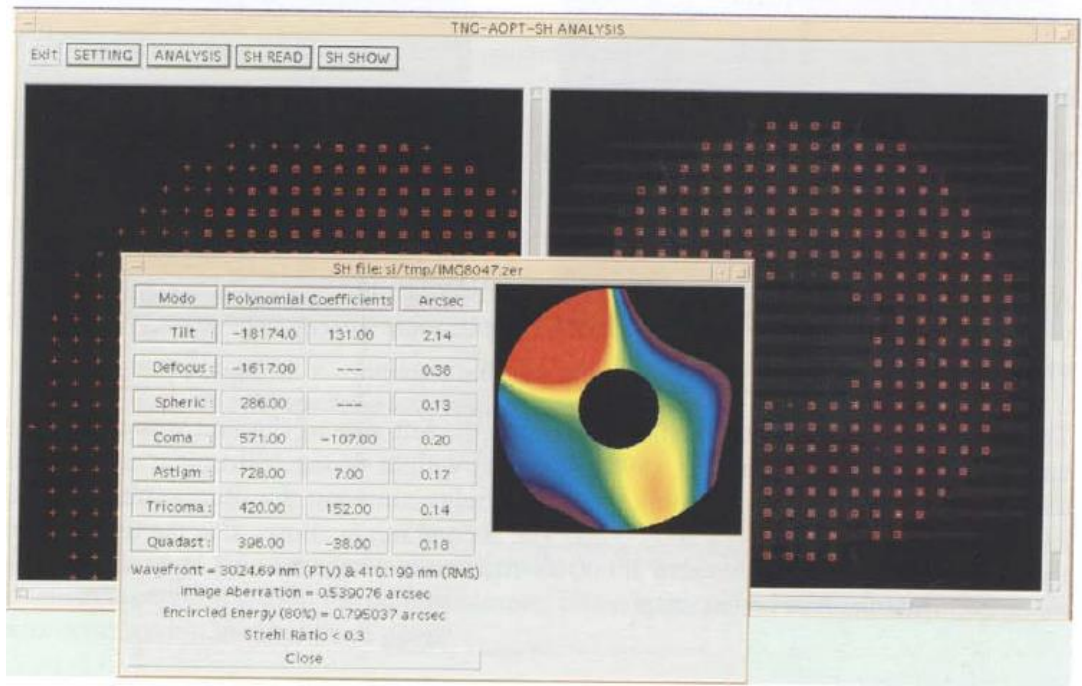

Fig. 7.4 - One of the first Shack-Hartman runs on the telescope optics.

The optimization made on the active optics system presently allows the operator to routinely reach an encircled energy of about 0.10-0.15 arcsec, close to the limits for the mirrors alone measured at Zeiss in 92.

An example of a current S-H analysis is shown in Fig. 7.5.

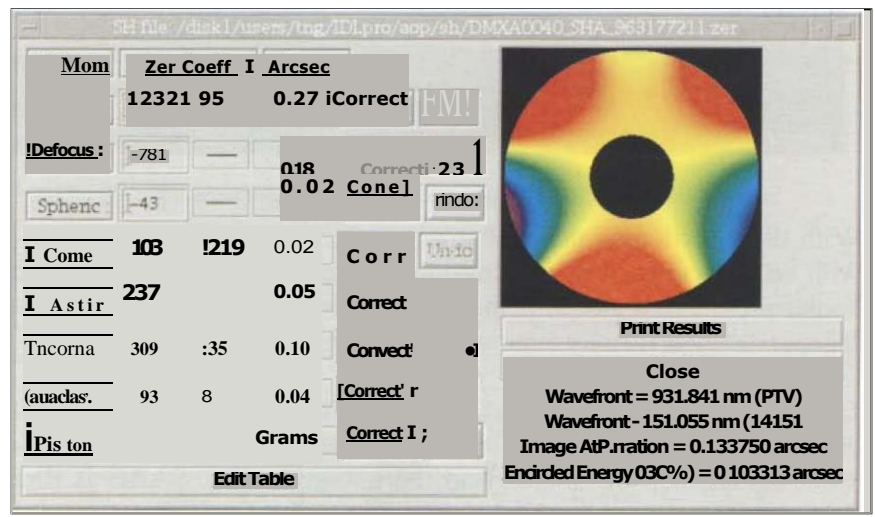

Fig. 7.5 - A typical S-H analysis.

TNG Scientific Dedication - F. Bortoletto. *et al.: The commissioning Phase*## 8. The instruments and the Nasmyth instrument interfaces

The first-light instrumentation plan for the Galileo telescope is described in Ref. 6 and 7. In Tab. 8.1 a summary of the instruments already mounted and in operation is presented.

With the exception of NICS, which is a NIR imager-spectrometer, all the spectroscopic instrumentation is mounted on the B-arm, while the imaging instrumentation is on the A-arm.

As shown in the schedule of Tab. 2.1, the commissioning of all the instrumentation, that was made possible also by the effort also of the resident TNG group, started at the end of 1998 and is now finished.

The Nasmyth-A instrument interface shown on the left side of Fig. 8.2 was completed and mounted during November 1998. The total weight of the instrument complex is about 2.5 tons. The interface holds the optical relay allowing the illumination of OIG and NICS in two modes: f-11 and f-32.2. The second mode allows also the insertion of the adaptive-optics module, also the OIG filter wheel system and light-shutter are mounted inside the instrument adapter.

**Tab. 8.1 — The TNG fist-light instrumentation plan**

| Instrument | Mode | Arm | Sensor | Spectr. range (microns) | Res. power ($\lambda/\Delta\lambda$) | Scale (arcsec/pix) | Field (arcminxarcmin) |
|---|---|---|---|---|---|---|---|
| OIG | Imaging | A | CCD | 0.32-1.1 | - | 0.07 | 5X5 |
| ARNICA | Imaging | A | NICMOS | 0.9-2.5 | - | 0.41 | 1.75X1.75 |
| NICS | Imaging | A | NICMOS | 0.9-2.5 | - | 0.25-0.13 | 4.3X4.3-2.2X2.2 |
| NICS | Spectroscopy | A | NICMOS | 0.9-2.5 | 300-1300 | (0.25) | 4.3 |
| ADOPT | Service | A | - | - | - | - | - |
| LRS | Imaging | B | CCD | 0.32-1.1 | - | 0.28 | 9.4X9.4 |
| LRS | Spectroscopy | B | CCD | 0.3-1.1 | 1500-3500 | - | 10 |
| SARG | spectroscopy | B | CCD | 0.35-1.1 | 19000-144000 | - | 8-30(arcsec) |

With the insertion of the NIR imager NICS, the NAS-A interface has been provided with an especially built rotating-joint in order to allow the transportation of the cryo-cooler pressurized gas. 1t is mounted in front of the interface on the rotation axis and independently driven by a motor and a tracking encoder placed in the derotator-A.

On the right panel of Fig. 8.2, the medium dispersion spectrograph LRS (now, *d.o.lo.re.s*) is shown directly mounted on the derotator axis and above the high dispersion spectrograph SARG mounted on the telescope arm. SARG is illuminated through a folding mirror placed before the slit holder of LRS.

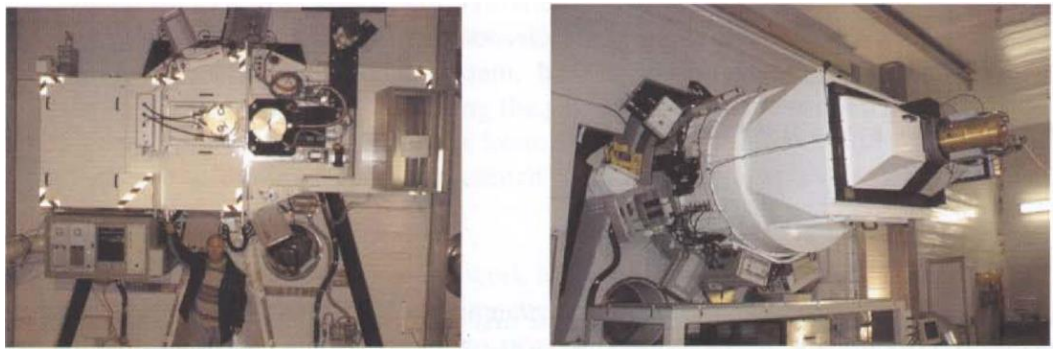

**Fig. 8.2** — The Nasmyth-A instrument interface after the assembly *(left panel),* and the LRS (now, *d.o.lo.re.s)* spectrograph mounted on the Nasmyth-B derotator *(right panel).*

## 9. Conclusions

A description of the main steps of work made on the Telescopio Nazionale Galileo and its instrumentation during the commissioning phase is presented in this paper. The work was done having in mind three goals:

- the telescope first light;

- the instrumentation first light;

- the start of the scientific operations.

The evolution of the foreseen schedule of operations, a comparison with the actual schedule, and some of the main difficulties we found and overcame are also presented.

All the data presented are coming from the first tests made on the system during the commissioning evolution. Nowadays the telescope accuracy is able to follow the external observing conditions up to half arcsecond of integrated PSF.

I would like to acknowledge all the people unintentionally not mentioned here, but well present during the effort, and, in particular, the small community of TNG personnel resident in La Palma for the reciprocal support during our years abroad.

**References**
*(TNG documents are available at :* **http://www.tng.iac.esidoc/publiclsw_comm/sw_comm.htrn1)**

1. TNG: Optics Commissioning
2. TNG: Software Commissioning
3. TNG: The Telescope First-Light
4. TNG: The Telescope User Manual
5. TNG: The Hydrostatic Bearing System
6. TNG: Instrument Plan, March 1992
7. TNG: Instrument Plan, Dec 1992
8. Bortoletto F., Bonoli C., D'Alessandro M., Ragazzoni R., Conconi P., Mancini D., Pucillo M. 1998, *"Commissioning of the Italian National Telescope Galileo",* Kona Conference on Astronomical Telescopes and Instrumentation, March 1998, SPIE Conf. Proc. Vol. 3352, p. 91
9. Bortoletto F. 1999, *"Commissioning of the Italian National Telescope Galileo",* Sheffield Conference on Medium Sized Telescopes and Instrumentation